\documentstyle[preprint,aps]{revtex}
\begin{document}
\draft
\title{The Energy Spectrum of the Membrane effective Model
for Quantum Black Holes}
\author{
C. O. Lousto\thanks{Permanent Address: Instituto de Astronom\'{\i}a y
F\'{\i}sica del Espacio, Casilla de Correo 67 - Sucursal 28, 1428 Buenos
Aires, Argentina. E-mail: lousto@iafe.uba.ar}}
\address{IFAE--Grupo de F\'{\i}sica Te\'orica, Universidad Aut\'onoma de
Barcelona, E-08193 Bellaterra (Barcelona), Spain. E-mail:
lousto@ifae.es}
\date{\today}
\maketitle
\begin{abstract}
We study first order fluctuations of a relativistic membrane in the curved
background of a black hole. The zeroth-order
solution corresponds to a spherical membrane tightly covering the event
horizon. We obtain a massive Klein-Gordon
equation for the fluctuations of the membrane's radial coordinate on the
2+1 dimensional world-volume.  We finally suggest that quantization of the
fluctuations can be related to black hole's mass quantization and the
corresponding entropy is computed. This entropy is proportional to the
membrane area and is related to the one-loop correction to the
thermodynamical entropy $A_H/4$.
With regards to the membrane model for describing effectively a quantum
black hole, we connect these results with previous work on critical
phenomena in black hole thermodynamics.
\end{abstract}

\paragraph*{1. Introduction} We will try
to find an effective (or phenomenological)
description of the black hole quantum degrees of freedom
in the spirit of the black hole {\it complementary principle}.
We shall consider the whole picture as described by
a fiducial observer (an observer at rest with respect to the black
hole as opposed to a
free falling observer). The principle of black hole complementarity
\cite{STU93,ST94} will ensure us not to have any logical contradiction
because of the non
existence of ``superobservers". For fiducial observers the black hole appears
well described classically by a membrane\cite{TPM86} sitting on the stretched
horizon (to regularize infinite redshift factors).

At quantum level several authors\cite{tH90,M94,L94} considered that the
relevant black hole's degrees of freedom to quantize were located
practically on the horizon surface. For the sake of definiteness we will
discuss the possibility of representing this black hole's degrees of freedom,
in an effective way, by a relativistic membrane. This model is an effective
one in the sense that it is not fundamental (nor renormalizable), but will be
eventually derivable (in the appropriate limit) from a consistent and
finite quantum theory of gravitation.

To fix ideas we can consider the effective system to study at low energies
as composed of the classical gravitational field (Einsteinian gravity, for
instance) and the relativistic membrane (plus, eventually, matter fields)
represented by the following Lagrangean density.
\begin{equation}
{\cal L}={\cal L}_{\text{grav}}+I_M\delta(X-X_M)+{\cal L_{\text{matter}}}
\label{a0}
\end{equation}
where the action of a relativistic bosonic membrane\cite{D89}
\begin{equation}
I_M=-{T\over 2}\int{d^3\xi\sqrt{-\gamma}\left\{g_{\mu\nu}(X)\gamma^{ij}
\partial_i{X^{\mu}}\partial_j{X^{\nu}}-1\right\}}~.
\label{a1}
\end{equation}
Here $T$ is the membrane tension. $\xi=(\tau,\sigma,\rho)$ are the
world-volume coordinates $(\mu,\nu=0,1,...,D-1.)$ and
$\partial_i =\partial /\partial\xi^i$ where the indices $i,j$ take
the values $0,1,2$.

We can think of the above action as emerging from a coarse-graining process
over very short distances from the black hole event horizon.
Our ignorance about these short distances is
hidden into a few phenomenological parameters such as the membrane tension,
$T$. One thus obtains, an effective lagrangian for the position of the
``stretched horizon'' (represented by collective variables) that, because of
the reparametrization covariance, is just the lagrangian of a relativistic
membrane in a curved background.

The advantage of this approach is that, since we have a concrete dynamical
model, we can compute the energy levels and
derive an associated entropy, and then check the consistence of this results
with what we already independently knew about quantum properties of black
holes, for finally to look for some new results.

It has already been observed by various
authors, e.g.~\cite{tH85,ST94},
that  short distances play an important role in
the counting of the number of states near the horizon, and that such a
counting gives a divergent result unless one introduces
a regulator such as a short distance
cutoff. Thus, we expect that also in our formulation
the separation between membrane
levels depend on such a cutoff, and we should investigate up to what
extent we can extract results which are independent of short
distance physics.

{}From action Eq.~(\ref{a1}) we can derive the classical equations of motion
\begin{equation}
g_{\mu\rho}(\Box_\gamma{X^{\rho}}+\Gamma^\rho_{k\lambda}(g)\partial_i{X^{k}}
\partial_j{X^{\lambda}}\gamma^{ij})=0~,
\label{a2}
\end{equation}
and the constraint
\begin{equation}
\gamma_{ij}=\partial_i{X^{\mu}}\partial_j{X^{\nu}}g_{\mu\nu}(X)
\label{a3}
\end{equation}
by varying with respect to $X^\mu$ and $\gamma_{\mu\nu}$ respectively.

We are interested in studying the membrane dynamics in the curved
background of a static
spherically symmetric black hole in $D=4$ dimensions
[In the following we work with Minkowskian signature, and
we  use Planck units, $\hbar=c=G=1$.]; its metric represented by

\begin{equation}
ds^2=-a(r)dt^2+b(r)^{-1}dr^2+r^2d\Omega^2~,~~
d\Omega^2=d\theta^2+\sin^2\theta d\varphi^2~.\label{a4}
\end{equation}

For a Reissner-Nordstr\"om black hole
\begin{equation}
a(r)=b(r)=1-2{M\over r}+{Q^2\over r^2}~,\label{a4b}
\end{equation}
and the classical event horizon is located at the radial coordinate
$r_+=M+\sqrt{M^2-Q^2}$.

We next conveniently choose the coordinate system according to our
description in terms of a
fiducial observer outside the black hole. In fact, it is from the point of
view of this observer that the black hole degrees of freedom
can be represented by a membrane. We have thus the following set
of coordinates
\begin{equation}
\label{a5}
X^0\dot=t~~,~~~X^1\dot=r~~,~~~X^2\dot=\theta~~,~~~X^3\dot=\varphi
\end{equation}

To study the semiclassical quantization of a membrane sitting on the
black hole event horizon we choose the, so called\cite{CT76},
{\it rest gauge} $X^0(\tau,\sigma,\rho)=\tau.$

Since we are thinking of studying the membrane fluctuations around a
classical solution having spherical symmetry, we will fix the gauge by taking
$X^2(\tau,\sigma,\rho)=\sigma$ and $X^3(\tau,\sigma,\rho)=\rho$
with $0\leq\sigma\leq\pi$ and $0\leq\rho\leq2\pi$.

\paragraph*{2. First Order Fluctuations}

The classical equations of motion in the
background of a Reissner-Nordstr\"om black hole have
solutions~\cite{M94} corresponding to membranes which approach the
background event horizon, $r=r_+$. If we try to quantize around this
solution, however, we find that the membrane level actually form a
continuum (see below), so that the number of states of the black hole
would diverge. This divergence is nothing but the divergence first found
 by 't~Hooft~\cite{tH85}, and can be regulated with a short
distance cutoff near the horizon. The physical origin of this cutoff,
in the membrane approach, is quite clear. The membrane is only an
effective theory, obtained performing a coarse graining over short
distances, that is, over distances on the order of a few Planck
lengths from the horizon (or a few string lengths, depending on what is
the fundamental theory), and it is not legitimate to extrapolate the
membrane equations of motion down to a distance smaller than a few
Planck length from the classical horizon.  Thus, we rather proceed as
follows. We assume that the equations of motion for $g_{\mu\nu}$, derived
form the lagrangian~(\ref{a0}), acquire corrections very
close to the horizon, in such a way that the membrane, instead of
approaching $r=r_+$ asymptotically, have an equilibrium position at
$X^1_M\doteq R_M=r_H+\epsilon^2/r_H$ with $\epsilon/r_H\ll 1,$
$(\epsilon\simeq$ a few Planck lengths).
Such corrections can produce a change in the dependence on $r$ of the
functions $a(r)$ and $b(r)$
which enter in the metric, see Eq.~(\ref{a4}). We do not present here any
explicit expression indicating how $a(r)$ and $b(r)$ are modified, since
it is clear that this is just a phenomenological way to take into account
short distance effects, and we will be interested only in those
results which have a rather general validity, and do not depend
on the specific properties of the modification introduced.

Let us then study the first order fluctuations in the only independent
coordinate left after our gauge choice
\begin{equation}
\label{a9}
X^{\mu}=\left(X^0(\tau), X^1(\tau,\sigma,\rho), \theta(\sigma),
\varphi(\rho)\right)~.
\end{equation}

{}From expressions~(\ref{a2}) and~(\ref{a3}) it is easy to see
that neglecting orders higher than the first, the equations of
motion~(\ref{a2}) become a
Klein-Gordon equation for the radial coordinate, $X^1$, in the $2+1$
world-volume of the membrane. In fact, $\gamma_{ij}$ has the following
components
\begin{eqnarray}
&&\gamma_{\tau\tau}=-a+b^{-1}(\dot X^1)^2~~,~~~\gamma_{\sigma\sigma}=
(X^1)^2+b^{-1}(X^{1\prime})^2~,\cr\cr
&&\gamma_{\sigma\tau}=\gamma_{\tau\sigma}=b^{-1}X^{1\prime}\dot X^1~~, ~~~
\gamma_{\sigma\rho}=\gamma_{\rho\sigma}=b^{-1}X^{1\prime}\tilde X^1~,\cr\cr
&&\gamma_{\tau\rho}=\gamma_{\rho\tau}=b^{-1}\dot X^1\tilde X^1~~,~~~
\gamma_{\rho\rho}=(X^1)^2\sin^2(X_2)+b^{-1}(\tilde X^1)^2~.
\label{a10}
\end{eqnarray}
where $\cdot=\partial_\tau$, $~'=\partial_\sigma$ and ~\~\ $=
\partial_\rho$~.

To first order in the fluctuations of $X^1$ around $R_M$ we are left with a
diagonal 3-metric
\begin{eqnarray}
\gamma^{(1)}_{ij}=\left(\matrix{-a&0&0\cr
                                0&R_M^2&0\cr
                                0&0&R_M^2\sin^2{(\theta)}\cr}\right)~.
\label{a11}
\end{eqnarray}

The field equations~(\ref{a2}) for the {\it r-component} in this case read
\begin{eqnarray}
&&{1\over b(R_M)}\left\{{-\partial^2_\tau \delta X^1\over a(R_M)}+
{1\over R_M^2\sin\theta}
\left[\partial^2_{\sigma}\delta X^1+\partial_{\rho}
{(\sin\theta\partial_{\rho}{\delta X^1})}\right]\right\}\cr\cr
&&-\left({2\over R_M^2}+{2a'(R_M)\over R_M a(R_M)}-{1\over4}\left({a'(R_M)
\over a(R_M)}\right)^2+{a''(R_M)\over 2 a(R_M)}\right)\delta X^1=
{2\over R_M}+{a'(R_M)\over 2a(R_M)}~,
\label{a12}
\end{eqnarray}
where $\delta X^1=X^1-R_M$.

For the {\it t-component}
\begin{equation}
\left({2\over R_M}+{a'(R_M)\over 2a(R_M)}\right)\partial_\tau \delta X^1=0~,
\label{a12b}
\end{equation}

for the {\it $\theta$-component}
\begin{equation}
\left({2\over R_M}+{a'(R_M)\over 2a(R_M)}\right)\partial_\sigma\delta X^1=0~,
\label{a12c}
\end{equation}

and for the {\it $\varphi$-component}
\begin{equation}
{1\over \sin\theta}\left({2\over R_M}+{a'(R_M)\over 2a(R_M)}\right)
\partial_{\rho} \delta X^1=0~.
\label{a12d}
\end{equation}

{}From Eqs.~(\ref{a12b})-(\ref{a12d}) we see that to have a non trivial
solution for $\delta X^1$ it must be
\begin{equation}
{2\over R_M}+{a'(R_M)\over 2a(R_M)}=0~.
\label{a13}
\end{equation}

This condition is in fact equivalent to the existence of a zeroth-order
solution representing
a spherical membrane at rest at $X^1=$ constant $=R_M$, and, precisely,
determines $R_M$ given the background metric~(\ref{a4}).
With Eq.~(\ref{a13}) fulfilled, Eqs.~(\ref{a12b})-(\ref{a12d}) are satisfied
automatically and we are left with Eq.~(\ref{a12}) as the only relevant
equation for $\delta X^1(\tau,\sigma,\rho)$.
It can be written as a massive 2+1 dimensional Klein-Gordon
equation on the sphere of radius $R_M$, (considering the proper time
$\tau '$ of an observer at rest at $R_M$, i. e. $\tau '=a(R_M)^{1/2}\tau$.)
Thus we have
\begin{equation}
\Box_{\tau '}\delta X^1-\mu^2\delta X^1=0~,
\label{a14}
\end{equation}
where
\begin{equation}
\mu^2=\left[{2\over R_M^2}+{2a'(R_M)\over R_M a(R_M)}-{1\over4}\left({a'(R_M)
\over a(R_M)}\right)^2+{a''(R_M)\over 2 a(R_M)}\right]b(R_M)
\label{a14b}
\end{equation}
is the effective mass squared of the first order fluctuations.

\paragraph*{3. Energy Spectrum and Mass Quantization}

Membrane theories are intrinsically non-linear and therefore much more
difficult to quantize than string theories. Thus, next, we will
perform a semiclassical quantization expanding around a
classical solution\cite{DIPSS88}. Working at first order and in a physical
gauge we obviate the problem of the constrained system quantization.

Let us decompose the solution to Eq.~(\ref{a14}) in spherical modes
\begin{equation}
\delta X^1(\tau,\sigma,\rho)=\sum_{l,m}\delta X^1_{lm}(\tau)
Y_{lm}(\sigma,\rho)~,
\label{a15}
\end{equation}
where $Y_{lm}(\theta,\varphi)$ are the usual spherical harmonics.

Plugging Eq.~(\ref{a15}) into Eq.~(\ref{a14}) we obtain the following
expression for the modes $\delta X^1_{lm}(\tau)$ of first order fluctuations
\begin{equation}
\ddot{\delta X^1_{lm}}+\omega_l^2\delta X^1_{lm}=0~,~~~
\omega_l^2=a(R_M)\left(\mu^2+\frac{l(l+1)}{R_M^2}\right)~.
\label{a16}
\end{equation}
This is just the classical equation of motion of an harmonic oscillator
with frequency $\omega_l$ for each mode labeled by $lm$.
As in any system with spherical symmetry $\omega_l$ does not depend
on $m$, the azimuthal angular momentum. This implies a degeneracy of
states of $d_l=2l+1$.

Upon quantization of these modes one obtains a discrete spectrum of
energies\cite{G95}
\begin{equation}\label{spectrum}
E_{nl}=(n+1/2)\omega_l~.
\end{equation}

In $\omega_l$, the mass term, $\mu$, is background dependent, but of less
relevance than the dependence on $\sqrt{a(R_M)}$ which is a non-trivial and
a largely model-independent feature. Physically, $a(R_M)^{1/2}$ is just the
redshift factor: the excitation of the membrane gives a contribution
to the energy at infinity of the black hole which is suppressed by a
factor equal to the redshift from the membrane location to
infinity. It is clear therefore that, if we remove the cutoff and send
naively $a(R_M)\rightarrow 0$, we get a continuum of energy levels.

One can interpret Eq.~(\ref{spectrum}) as indicative of a discrete spectrum
of black hole masses. The spacing between two neighbour levels is given by
\begin{equation}
\Delta E_n={E_{nl}\over n+1/2}~,~~\text{for fixed $l$.}
\label{a17}
\end{equation}
And
\begin{equation}
\Delta E_l={2a(R_M)(n+1/2)^2(l+1)\over R_M^2(E_{nl}+E_{n,l+1})}
\simeq{E_{nl}\over l}~,~~\text{for fixed $n$.}
\label{a18}
\end{equation}

We observe that both level separations are of the same order and tend to zero
as $l$ or $n$ goes to infinity (continuum limit). [In the brick wall model
\cite{tH85} one would have $\Delta E\sim\sqrt{a(R_M)}/R_M\sim\epsilon/M^2$.]
This result, interpreted as a black hole mass quantization gives much thinner
level separation than earlier estimates\cite{B74,M90,K86,G93,P93,L95}.
Suppose that  the black hole mass $E_{nl}$ is
large but not astronomically large compared to the Planck mass
$M_{\text{Pl}}$, say $E_{nl}\sim 10^3 M_{\text{Pl}}$;
then the separation $\Delta E_{nl}\sim 10^{-6}M_{\text{Pl}}$
is  small compared to $M_{\text{Pl}}$, but still
not small enough to allow a particle with ordinary energy, say a few GeV,
to be absorbed or emitted by the black hole.
Thus, the emission line-spectrum would still be
significantly different from the semiclassical (continuum) result,
even for black holes with  mass $M\gg M_{\text{Pl}}$, for which we
would rather expect that the semiclassical approximation works well.

\paragraph*{4. Entropy}

The contribution to the entropy of the energy levels~(\ref{spectrum}) can
be obtained from the partition function
\begin{equation}\label{part}
{\cal Z}=\prod_{l=0}^\infty\prod_{m=-l}^l\sum_{n=0}^\infty e^{-\beta
E_{nl}}~,
\end{equation}
where $\beta=1/T$ and the sum over $n$ is essentially a geometric series.
Then,
\begin{equation}\label{logpart}
\ln{\cal Z}=-\sum_{l=0}^\infty(2l+1)\ln{\left[e^{-\beta\omega_l/2}\over
1-e^{-\beta\omega_l}\right]}
\end{equation}
[Note that we have taken the sum over $n$ up to infinity, although the
expression~(\ref{spectrum}) for $E_{nl}$ was derived in an approximation
that makes it valid in the regime $E_{nl}\ll M$. However the big $n$
contribution to ${\cal Z}$ is negligible.]

The mean energy per mode is
\begin{equation}\label{meanenergy}
<E>=\partial_\beta\ln{\cal Z}=\sum_{l=0}^\infty(2l+1)\left[
{\beta\omega_l\over e^{\beta\omega_l}-1}+{\beta\omega_l\over 2}
\right]_{\beta=T_{BH}^{-1}}~,
\end{equation}
wherefrom we see that the spectrum of radiation of this system will be
essentially planckian. [The $\beta\omega_l/2$ term is the usual vacuum
polarization contribution to be renormalized away upon operator ordering.]

{}From the expression for the entropy
$S=\left[\ln{\cal Z}-\beta\partial_\beta
\ln{\cal Z}\right]_{\beta=T_{BH}^{-1}}$ we obtain
\begin{equation}\label{entropy}
S_M=-\sum_{l=0}^\infty(2l+1)\left[\ln\left(1-e^{-\beta\omega_l}\right)
-{\beta\omega_l\over e^{\beta\omega_l}-1}\right]_{\beta=T_{BH}^{-1}}~.
\end{equation}

To perform the sum above we approximate it by an integral over $l$. We then
change to the variable $x=\beta\omega_l$ and obtain
\begin{equation}\label{intentropy}
S_M=-{2R_M^2\over\beta^2a(R_M)}\int_{\beta\omega_0}^\infty dx x
\left\{\ln\left(1-e^{-x}\right)-{x\over e^{x}-1}\right\}~
\Bigg|_{\beta=T_{BH}^{-1}}~.
\end{equation}

Since $\beta\omega_0=\beta\sqrt{a(R_M)}\mu\ll1$ the integral gives
\begin{equation}\label{entrofinal}
S_M=6\zeta(3)\left({T_{BH}^2\over a(R_M)}\right)R_M^2+
(\mu R_M)^2\left(\ln\left[{\mu\sqrt{a(R_M)}\over T_{BH}}\right]
+3/2\right)~,~~\zeta(3)=1.20206~.
\end{equation}

The first term in this expression is the leading term. It is proportional to
the membrane (or stretched horizon) area and to the square of the local
temperature. The logarithmic term here appears due to the $l=0$ mode
contribution to the energy levels. It can be taken as a sample of what we
would expect from non-leading corrections to the one-loop entropy.
[Note that in Eq.~(\ref{entrofinal}) the same dependence on the redshift
factor would appear had we considered observations at $R_M$ instead of at
infinity, since $\beta E_{nl}|_\infty=\beta E_{nl}|_{R_M}$.]

So far, in Eq.~(\ref{entrofinal}) does not appears explicitly any cut-off.
If we introduce 't Hooft's brick wall\cite{tH85}, $R_M=2M+\epsilon^2/(2M)$,
we would obtain $S_M\sim C_1 M^2/\epsilon^2+C_2\ln(M/\epsilon)$. The leading
term corresponds to the one obtained by other methods such as the study of
four - dimensional quantum fields in the Schwarzschild
background\cite{tH85,FN93,S93,CW94,KS94,D94,F94,ML95}.
[We note that this result is non-trivial in our case, since it was
obtained from the study of a Klein--Gordon field in 2+1 dimensions.]

For further comparison it is interesting
to recall here that in Ref. \cite{LS88} it was studied the
back reaction effect of the Hawking radiation on the background Schwarzschild
black hole metric. It was thus obtained a correction to the
Bekenstein--Hawking temperature of the following form
\begin{equation}\label{temperature}
T_{BH}={1\over 8\pi M}\left[1+{\alpha\over M^2}\right]
\end{equation}
where $\alpha$ is a constant depending on the field content of the Hawking
radiation. From the above equation we obtain for the entropy
\begin{equation}\label{entropia}
S_{BH}=\int {dM\over T}=4\pi M^2-8\pi\alpha\ln M~.
\end{equation}

We thus see that the effect of the one-loop corrections
to the entropy found above
can be interpreted, respectively, as producing a renormalization of
the gravitational constant $G$ and the higher order
coupling constants (Back reaction terms in our case or
purely vacuum polarization ones in the case studied in
Ref. \cite{DLM95}).

\paragraph*{5. Discussion}

In this paper we have taken the relativistic membrane as a phenomenological
model to describe the quantum degrees of freedom of a black hole. The main
results we have obtained are the energy spectrum, i.e. Eq.~(\ref{spectrum})
and its associated entropy, i.e.
Eq.~(\ref{entrofinal}). The main approximations
in arriving to these results have been the linearization of the system (and
considering $l$ as a continuum variable in the computation of the entropy).
It is clear from the lagrangian formulation~(\ref{a0}) that the membrane
contributions are perturbative with respect to the classical (or tree-level)
expressions. Then, in particular, the total entropy will be given by
$S_{BH}=A_H/4+S_M$ and thus, the membrane corrections can be associated to
the one-loop level. Within this approach we did not address the questions
of the stability of the classical membrane solution at a given radius,
$R_M$, near the black hole event
horizon as well as the origin of the Hawking radiation at a black hole
temperature $T_{BH}$, but rather consider them as external parameters
provided by the curved background. Also covariant higher order fluctuations
and careful quantization of the constrained system should be considered.
All these points deserve a particular study and will indeed be the line of
research to follow in obtaining a completely self consistent picture of
which the present work must be considered only a first step.

In Refs.\cite{L93,L94,L95,CL95} it was shown strong
evidence for that four-dimensional,
rotating and charged black holes undergo critical
phenomena. Under critical conditions their characteristic behaviour is as if
they had an effective dimension equal to two. In fact, critical
exponents of correlation functions
 and quite general arguments coming from the Renormalization Group
theory assign an effective (spatial) dimension, $d=2$, to the system.
Also, by comparison of the
black hole critical exponents with those of the Gaussian model in
$d$-dimensions, complete agreement is only found for $d=2$.
We can think of  Eq.~(\ref{a14}) as coming from an action for the ``field"
$\delta X^1(\tau,\sigma,\rho)$
\begin{eqnarray}
I_M(\delta X^1)&=&-{T\over2}\int d\tau d\theta d\varphi R_M^2a(R_M)^{-1/2}
\sin\theta\times\cr
&&\left\{-{(\partial_{\tau}{\delta X^1})^2\over a(R_M)}
+{1\over R_M^2}\left[(\partial_{\theta}{\delta X^1})^2+
{1\over\sin\theta}
(\partial_{\varphi}{\delta X^1})^2\right]-\mu^2(\delta X^1)^2\right\}~.
\label{a14c}
\end{eqnarray}
which generates essentially a Gaussian integral and we can~\cite{L95},
then, reproduce in the membrane approach
all the black hole critical exponents since
$\delta X^1(\tau,\sigma,\rho)$ is now a Gaussian field living on a spatially
two-dimensional sphere of radius $R_M$.

\begin{acknowledgments}
The author is very much indebted to M.Maggiore and C.Gundlach for
enlighting discussions and help with the computations.
This work was partially supported by the Directorate General for
Science, Research and Development of the Commission of the European
Community. C.O.L was supported by the Direcci\'on General de
Investigaci\'on Cient\'\i fica y T\'ecnica of the Ministerio de
Educaci\'on y Ciencia de Espa\~na and CICYT AEN 93-0474.
\end{acknowledgments}


\begin{references}

\bibitem{STU93} L.Susskind, L.Thorlacius and J.Uglum,
Phys. Rev. D 48 (1993) 3743.\par

\bibitem{ST94} L.Susskind and L.Thorlacius, Phys. Rev. D 49 (1994) 966.\par

\bibitem{TPM86} K.S.Thorne, R.H.Price and D.A.Macdonald (Eds.),
Black holes: The membrane Paradigm (Yale Univ. Press, New Heaven, 1986).\par

\bibitem{tH90} G.'t Hooft, Nucl. Phys. B, 335 (1990) 188.\par

\bibitem{M94} M.Maggiore, Nucl. Phys. B, 429 (1994) 205.\par

\bibitem{L94} C. O. Lousto, in {\it Cosmology and Particle Physics},
V. De Sabbata and H. Tso-Hsiu Eds., NATO ASI series Vol. C427,
p 183-192, Kluwer Academic Publishers, Dordrecht (1994).\par

\bibitem{D89} M.J.Duff, Class. Quantum Grav., 6 (1989) 1577.\par

\bibitem{tH85} G.'t Hooft, Nucl. Phys. B, 256, (1985) 727.\par

\bibitem{CT76} P.A.Collins and R.W.Tucker, Nucl. Phys. B, 112 (1976) 150.\par

\bibitem{DIPSS88} M.J.Duff, T.Inami, C.N.Pope, E. Sezgin and K.S.Stelle,
Nucl. Phys. B, 297 (1988) 515.\par

\bibitem{G95} C.Gundlach, preprint LAEFF-95-2, gr-qc/9501008.\par

\bibitem{B74} J. Bekenstein, Lett. Nuovo Cimento, 11 (1974) 467.\par

\bibitem{M90} V.F. Mukhanov, in {``\em Complexity, Entropy and the Physics
of Information''}, Ed. W. Zurek, Addison-Wesley, Redwood City, p 47
(1990); JETP Lett. 44 (1986) 63.\par

\bibitem{K86}  Ya. I. Kogan, JETP Lett. 44 (1986) 267.\par

\bibitem{G93} J. Garc\'\i a - Bellido,
preprint SU-ITP-93-4, hepth/9302127.\par

\bibitem{P93} Y. Peleg, preprint BRX-TH-350, hep-th/9307057.\par

\bibitem{L95} C.O.Lousto, Phys.Rev., D51, (1995) 1733.\par

\bibitem{FN93} V.P.Frolov and I.D.Novikov, Phys. Rev.
D, 48, (1993) 4545.\par

\bibitem{S93} M.Srednicki, Phys. Rev. Lett. 71, (1993) 666.\par

\bibitem{CW94} C.Callan and F.Wilczek, Phys. Lett. B, 333, (1994) 55.\par

\bibitem{KS94} D.Kabat and M.J.Strassler, Phys. Lett. B, 329, (1994) 46.\par

\bibitem{D94} J.S.Dowker, Class. Quantum Grav., 11, (1994) L55.\par

\bibitem{F94} V.P.Frolov, preprint ALBERTA-THY-22-94, gr-qc/9406037.\par

\bibitem{ML95} R.M\"uller and C.O.Lousto, ``Entanglement entropy in
curved spacetimes with event horizon", preprint UAB-FT-362, (1995).\par

\bibitem{LS88} C.O.Lousto and N.S\'anchez, Phys. Lett.
B, 212, (1988) 411.\par

\bibitem{DLM95} J.-G.Demers, R.Lafrance and R.Myers, preprint McGill/95-06,
gr-qc/9503003.

\bibitem{L93} C.O.Lousto, Nucl. Phys. B, 410 (1993) 155.\par

\bibitem{CL95} C.O.Lousto, J.Gen.Rel.Grav., Vol.27, (1995) 121.\par

\end{references}
\end{document}